\documentclass[preprint,12pt,authoryear]{elsarticle}
\usepackage{threeparttable}
\usepackage{amsmath,amssymb,amsthm}
\usepackage{multirow}
\usepackage{booktabs}
\usepackage{rotating}
\usepackage{color}

\begin{document}

\begin{frontmatter}
\title{H-relative error estimation approach for multiplicative regression model with random effect}

\author[a]{Zhanfeng Wang\corref{cor1}}
\author[a]{Zhuojian Chen}
\author[a]{Zimu Chen}
\address[a]{Department of Statistics and Finance, Management School, University of Science and Technology of China, Hefei, China.}
\cortext[cor1]{Department of Statistics and Finance, Management School, University of Science and Technology of China, Hefei, China. (Email: zfw@ustc.edu.cn).}

\begin{abstract}
Relative error approaches are more of concern compared to absolute error ones such as the least square and least absolute deviation, when it needs scale invariant of output variable, for example with analyzing stock and survival data. An h-relative error estimation method via the h-likelihood is developed to avoid heavy and intractable integration for a multiplicative regression model with random effect. Statistical properties of the parameters and random effect in the model are studied. To estimate the parameters, we propose an h-relative error computation procedure. Numerical studies including simulation and real examples show the proposed method performs well.
\end{abstract}

\end{frontmatter}

\section{Introduction}

	In regression analysis, the least squares (LS) and least absolute deviation (LAD) are the most commonly used criteria based on absolute errors \citep{stigler1981gauss, portnoy1997gaussian}.
However, relative error methods are more of concern when it needs scale invariant of response variable (output) such as analyzing stock price and survival data
\citep{narula1977prediction, makridakis1984forecasting, khoshgoftaar1992predicting, ye2007price, park1998relative, Chen:2010ic, zhang2013local, wang2015change, liu2016group, chen2016least}.
For example, based on multiplicative regression models, \cite{Chen:2010ic} proposed a least absolute relative error (LARE) method, and \cite{wang2015change} developed a relative error change-point estimation approach. The LARE criterion was used to construct a local least absolute relative error estimation for a partially linear multiplicative model \citep{zhang2013local}.
For a more flexible model, multiplicative single index model, \cite{2016arXiv1609.01553} showed a two-step estimation procedure to estimate the parameter and unknown link function with respect to relative errors.

 However, the preceding relative error methods do not take a random effect account in their studied models. To the best of our knowledge, most of random effect approaches in literature are built on the absolute error or likelihood methods. It is much desired to study a relative error method for random effect models. When relative errors are of concern, the response variable generally is positive. Similar to the multiplicative regression model in \cite{chen2016least}, we construct an h-relative error approach based on the following model
\begin{equation}\label{rmodel}
Y=\exp(X^ {T}\beta+\nu)\epsilon,
\end{equation}	
where $Y$ is the response variable, $X$ is the $p$-vector of explanatory variables with the first component being 1 (intercept), $\beta$ is the corresponding $p$-vector of regression parameters with the first component being the intercept, $\nu$ is the random effect and $\epsilon$ is the error term which is strictly positive.
In model (\ref{rmodel}), when $\nu=0$, \cite{chen2016least} proposed a least product relative error criterion which possesses some merits: smooth, convex and so on. They stated that under
the error $\epsilon$ with the density
\begin{equation}\label{density}
f(t)=c\exp\{-t-1/t-\log(t)+2\}I(t>0),
\end{equation}
the parameter estimator is asymptotically efficient, where $c$ is a normalizing constant.

The density (\ref{density}) is employed to develop a computation algorithm for the parameter estimation in this paper. The likelihood principle \citep{birnbaum1962foundations} states that marginal likelihood of $\beta$ carries all the information in the data about the fixed parameters $\beta$,
so that the marginal likelihood should be used for inferences about $\beta$.
However, in general the marginal likelihood requires intractable integration which is usually hard to obtain a precise result. One method to obtain the marginal maximum likelihood (ML) estimator for $\beta$ is the expectation-maximization (EM) algorithm in \cite{dempster1977maximum}.
But, the EM algorithm is ofen numerically slow to converge and other simulation methods, such as Monte Carlo EM \citep{vaida2004mixed} and Gibbs sampling \citep{karim1992generalized} are computationally intensive. Instead, numerical integration using Gauss-Hermite quadrature \citep{crouch1990evaluation} could be directly applied to obtain the ML estimators, but this also becomes computationally heavier as number of random components increases.

To avoid heavy and intractable integration, in this paper, we employ the h-likelihood method \citep{lee1996hierarchical, lee2001hierarchical, lee2005likelihood} to build an h-relative error method to estimate the parameter $\beta$ and the random effect $\nu$. The proposed method also inherits scale invariance and less sensitive to outliers \citep{Chen:2010ic}. We develop asymptotic properties of $\beta$ and $\nu$, such as
consistence and normality. A computation algorithm is proposed to estimate the parameters via the h-relative errors. Numerical studies show the proposed method has better performance of the parameter estimation  compared to the traditional linear mixed model.

The rest of this paper is organized as follows. Section 2 introduces relative errors, parameter estimation for model (\ref{rmodel}), and provides their statistical properties.
Numeric studies including simulation and real examples are in Section 3. All of proofs of the theorems are presented in Appendix.

\section{Methodologies}
\subsection{H-likelihood with relative errors}

Throughout this paper we mean by $c$ the positive constant independent of $n$, which may take different values in different formulae or even in different parts of one and the same inequality.

Suppose observation samples $(Y_{ij}, X_{ij}),~i=1,...,K, j=1,...,n_{i}$ are randomly generated from
model (\ref{rmodel}) with repeatedly measured responses
\begin{equation}\label{rmodel-obs}
Y_{ij}=\exp(X_{ij}^ {\top}\beta+\nu_i)\epsilon_{ij}, ~i=1,...,K, j=1,...,n_{i}.
\end{equation}	
It follows that $Y_{i}=(Y_{i1},Y_{i2},...Y_{in_{i}})^T$ is the response for the $i$th unit $(i=1,...,K)$ and $\nu_{i}$ is the corresponding unobserved random effect.
Let the total sample size $n=\sum_{i=1}^{K}n_{i}$. Without loss of generality, let $\nu_i\sim N(0,\sigma^2)$ in this paper. The proposed methods can be extended to models with random effect having other distributions. Since $\epsilon_{ij}$ has the density from function (\ref{density}), conditional density
function $f(Y_{ij} | \nu_{i}; \theta)$ satisfies
\begin{equation}
\log(f(Y_{ij} | \nu_{i}; \beta))\equiv -Y_{ij}\exp(-X_{ij}^T\beta-\nu_{i})-\exp(X_{ij}^T\beta+\nu_{i})/Y_{ij}+c,\nonumber
\end{equation}
which suggests that
\begin{equation}\label{lpre}
-\log(f(Y_{ij} | \nu_{i}; \beta))=\Big|\frac{Y_{ij}-\exp(X_{ij}^ {\top}\beta+\nu_i)}{Y_{ij}}\Big |\times\Big |\frac{Y_{ij}-\exp(X_{ij}^ {\top}\beta+\nu_i)}{\exp(X_{ij}^ {\top}\beta+\nu_i)}\Big|+c.
\end{equation}
It shows that
\begin{equation}
\Big|\frac{Y_{ij}-\exp(X_{ij}^ {\top}\beta+\nu_i)}{Y_{ij}}\Big |~~{\mbox{and}}~~\Big |\frac{Y_{ij}-\exp(X_{ij}^ {\top}\beta+\nu_i)}{\exp(X_{ij}^ {\top}\beta+\nu_i)}\Big|\nonumber
\end{equation}
are two types of relative error: one is the error relative to the target and the other is the error relative to the predictor of the target.
It shows that (\ref{lpre}) consists of these relative errors, which leads to a relative error estimation criterion with respect to likelihood technique.

For convenience of notations, let $\theta=(\beta^T,\sigma^{2})^T$, $\nu=(\nu_{1},\nu_{2},...,\nu_{K})^T$ and $Y=(Y_{1}^T,Y_{2}^T,...,Y_{K}^T)^T$, where $\beta$ denotes the location parameter and $\sigma^2$ denotes the dispersion parameter. For model (\ref{rmodel-obs}), a log h-likelihood on the parameters is defined as
\begin{equation}\label{h-likelihood}
H\{\theta,\nu;Y\}=\sum_{i=1}^{K}h_{i}\{\theta,\nu_{i};Y_{i}\}=
\sum_{i=1}^{K}\Big(l_{1i}\{\theta,v_{i};Y_{i}\}+l_{2i}\{\theta,v_{i}\}\Big),
\end{equation}
where
\begin{align}
&l_{1i}\{\theta,v_{i};Y_{i}\}=\sum_{j=1}^{n_{i}}\log f(Y_{ij} | \nu_{i}; \beta),\nonumber\\
&l_{2i}\{\theta,v_{i}\}=\log f(\nu_i;\sigma^2)=-\frac{1}{2}\log(\sigma^2)-\frac{1}{2\sigma^2}\nu_{i}^2.\nonumber
\end{align}
It easily shows that $\sum_{i=1}^{K}l_{1i}\{\theta,v_{i};Y_{i}\}$ becomes the least product relative error criterion (LPRE) in
Chen et al. (2016).

To introduce the connection between data generation and parameter estimation, this paper also considers the extended likelihood framework \citep{lee2005likelihood}
\begin{equation}
L(\theta,\nu;y,\nu) \equiv f_{\theta}(\nu,y) = f_{\theta}(\nu)f_{\theta}(y|\nu)=f_{\theta}(y)f_{\theta}(\nu | y).\nonumber
\end{equation}
Then $H\{\theta,\nu;Y\}$ can be rewritten as
\begin{equation}\label{another-h}
H\{\theta,\nu;Y\}=m(\theta,Y)+l(\theta,\nu|y),
\end{equation}
where
\begin{align}
& m(\theta,Y)=\log L(\theta,Y)=\sum_{i=1}^{K}\log \int e^{h_i\{\theta,\nu_i;Y_i\}}d\nu_{i}=\sum_{i=1}^{K}m_i(\theta,Y),\label{marginal-likelihood}\\
& l(\theta,\nu|y)=\sum_{i=1}^{K}l_{i}(\theta,\nu_i|y)=\sum_{i=1}^{K} \log\frac{f(Y_{i} | \nu_{i}; \beta)f(\nu_i;\sigma^2)}{\int e^{h_{i}(\theta,\nu_{i};Y_{i})}d\nu_{i}}.\label{posterior}
 \end{align}
It follows that $m(\theta,Y)$ is a marginal log-likelihood with respect to $\theta$ and $l(\theta,\nu|y)$ is a conditional density function of $(\theta,\nu)$ for given data $Y$.

Estimators of the random effect $\nu_i$ and parameter $\theta$ are obtained by maximizing the log h-likelihood (\ref{h-likelihood}) or the log extended likelihood (\ref{another-h}), that is why we call h-relative error estimation approach for multiplicative regression model with random effect.



\subsection{Inference on random effects}

 We firstly treat $\theta$ as known. Inference on $\nu_{i}$ with an estimate of $\theta$ will be discussed later in this section.
We know estimation of $\nu_{i}$ only involves information from the $i${th} subject when $\theta$ is fixed.
From the h-likelihood method \citep{lee1996hierarchical}, it gives an estimator of $\nu_{i}$, saying $\hat\nu_i$, by solving the equation
$$h_{i}^{(1)}\{\theta,\nu_{i};Y_{i} \}= 0,$$
where $h_{i}^{(k)}\{\theta,\nu_{i};Y_{i} \}={\partial^k h_i}/{\partial \nu_{i}^k}$, $k=1,2,..,6$.

For $r=r(\nu)$, the quantity $\delta = E(r|y)$ is the best unbiased predictor for the $r$ in the sense that $E(\delta)=E_{y}{E(r|y)}=E(r)$. And it has the minimum mean-square error of prediction with respect to $E(\delta-r)'P(\delta-r)$ for any positive define matrix $P$.

Under appropriate conditions, we show that $\hat{\nu}$ converges to $E(\nu|y)$ presented in the following theorem, which proof is in Appendix.
\newtheorem{theorem}{Theorem}
\begin{theorem}
Under conditions $A_{1}$ in Appendix hold, we have
\begin{align}
&\hat{\nu_{i}}=E(\nu_{i} | Y)+O_p(\frac{1}{n}),
&Var(\nu_{i} | Y) = D_{i}^{*-1}\{1+O_{p}(n^{-1})\},\nonumber
\end{align}
where  $D_{i}^*=-{\partial^{2} H }/{\partial \nu_{i}^2} |_{\nu_{i}=\hat{\nu}_{i}}$.
\end{theorem}

For given $\theta$, the Laplace approximations of the expressions  (\ref{h-likelihood}) and (\ref{another-h}) with respect to the random effect are
\begin{align}
&\hat{H}\propto h_i\{\theta,\hat{\nu}_i;Y_i\}-\frac{1}{2}(\hat{\nu}_{i}-\nu_{i})'D_{i}^*(\hat{\nu}_{i}-\nu_{i}),\label{appro-H}\\
&\hat{H}\propto {m}_i(\theta;Y_i)+\hat{l}_i(\theta;\nu_{i} | Y_i),\label{appro-H-1}
\end{align}
where $\hat{H}$ and $\hat l_i$ are separately Taylor expansions of $H$ and $l_i$ with respect to $\nu_i$ at point $\nu_i=\hat{\nu_{i}}$.
Since $m_i$ and $h_i\{\theta,\hat{\nu}_{i};Y_i\}$ do not depend on $\nu_{i}$, they can be ignored when the distribution of $\nu_{i}|Y$ is computed.
Thence, (\ref{appro-H}) and  (\ref{appro-H-1}) imply that
\begin{equation}
\nu_{i} | Y \propto N(\hat{\nu}_{i},D_{i}^{*^{-1}}).\nonumber
\end{equation}
It follows that $N(\hat{\nu}_{i},D_{i}^{*^{-1}})$ is a reasonable approximation distribution of $\nu_i|Y$.
Easily we show that $D_{i}^{*-1}$ has order of $O_{p}(n_{i}^{-1})$ under the assumption $A_{1}$.

Under unknown $\theta$, following \cite{paik2015frequentist}, let $(\hat{\theta},\hat{\nu})$ be a solution of

\begin{align}
& \left (
 \begin{matrix}
   \frac{\partial}{\partial \theta}m(\theta;Y)   \\
   W\{\theta,\nu;Y\}
  \end{matrix}
  \right )  =0,
\end{align}
where $W\{\theta,\nu;Y\}=(h_{1}^{(1)}\{\theta,\nu_{1};Y_{1}\},...,h_{K}^{(1)}\{\theta,\nu_{K};Y_{K}\})^\top$. For a realized value $\nu_{0i}$, we can have the next theorem, which is similar to \cite{paik2015frequentist}.

\begin{theorem}
Under Conditions $A_{1}$ and $A_{2}$ holds,$\sqrt{n_{i}}(\hat{\nu}_{i}-\nu_{0i})$ converges in distribution to normal with mean $0$ and variance
\begin{align}
& I(\theta,\nu_{0i})^{-1}+n_{i}I(\theta,\nu_{0i})^{-1}B_{21i}A_{11}^{-1}Var \Big[ \frac{\partial}{\partial \theta}m_i(\theta;Y_i) | \nu_{i} = \nu_{0i} \Big] A_{11}^{-1} B_{21i}^{\top} I(\theta,\nu_{0i})^{-1} \nonumber\\
& -2n_{i}I(\theta, \nu_{0i})^{-1}B_{21i}^{\top}A_{11}^{-1}Cov \Big[ \frac{\partial}{\partial \theta} m_i(\theta;Y_i) | \nu_{i} ,h_{i}^{(1)}\{ \theta,\nu_{i};Y_{i} \} | \nu_{i} = \nu_{0i} \Big],\nonumber
\end{align}
where $A_{11}=E \{ -\frac{\partial^2}{\partial \theta \partial \theta^{\top}}m_i(\theta;Y_i) \}$ , $B_{21i} = \frac{1}{n_{i}} E \{ \frac{\partial}{\partial \theta^{\top}}h_{i}^{(1)} \{ \theta, \nu_{i};Y_{i} \} | \nu_{i} = \nu_{0i} \}$ and $I(\theta,\nu_{0i})= \frac{1}{n_{i}^2}E \Big[ -h_{i}^{(2)} \{ \theta ,\nu_{i} ;Y_{i} \} | \nu_{i} = \nu_{0i} \Big]$.
\end{theorem}
	
\subsection{Inference on location parameters}

Naturally, one way to estimate the parameter $\beta$  is to maximize the marginal log-likelihood
$m(\theta,Y)$, but the integration involved in  $m(\theta,Y)$ is intractable. Following \cite{paik2015frequentist}, we use a Laplace approximation method to compute the integration.
For convenience of notations, let the Laplace approximation of
a function $l(\alpha)$ on $\alpha$, be
\begin{equation}\label{laplace}
p_{\alpha}(l)=\Big[l-\frac{1}{2}\log \, \det\{D(l,\alpha)/(2\pi)\}\Big]\Big|_{\alpha=\tilde{\alpha}}
\end{equation}
where $D(l,\alpha)=-\partial^2 l /\partial \alpha^2$ and $\tilde{\alpha}$ is one solution of the equation $\partial l /\partial \alpha =0$.

Next we show that $p_{\nu}(H)$ is a reasonable approximation to $m(\theta,Y)$.
From \cite{tierney1986accurate},  we show that
\begin{equation}
\exp\{m_i(\theta;Y_i)\}=\int e^{h_{i}\{\theta,\nu_{i};Y_{i}\}}d\nu_{i}=e^{h_{i}\{\theta,\hat{\nu}_{i};Y_{i}\}}\sqrt{2\pi}
\tau_{i} n_{i}^{-\frac{1}{2}}[1-C_{n_{i}}\{\theta,\hat{\nu}_{i}\}]+O(n_{i}^{-2}),
\end{equation}
where
\begin{align}
&\tau_{i}^2 = -[h_{i}^{(2)}\{\theta,\hat{\nu}_{i};Y_{i}\}]^{-1},\nonumber\\
&C_{n_{i}}\{\theta,\hat{\nu}_{i}\}=J_{1i}\{\theta,\hat{\nu}_{i};Y_{i}\}/{(8n_{i})}-{5}
J_{2i}\{\theta,\hat{\nu}_{i};Y_{i}\}/{(24n_{i})},\nonumber\\
&J_{1i}\{\theta,\hat{\nu}_{i};Y_{i}\}=-{h_{i}^{(4)}\{\theta,\hat{\nu}_{i};Y_{i}\}}/{[h_{i}^{(2)}\{\theta,\hat{\nu}_{i};Y_{i}\}]^2},\nonumber\\
&J_{2i}\{\theta,\hat{\nu}_{i};Y_{i}\}=-{[h_{i}^{(3)}\{\theta,\hat{\nu}_{i};Y_{i}\}]^2}/{[h_{i}^{(2)}\{\theta,\hat{\nu}_{i};Y_{i}\}]^3}.\nonumber
\end{align}
For model (\ref{rmodel-obs}), it easily shows that $C_{n_{i}}$, $J_{1i}\{\theta,\hat{\nu}_{i};Y_{i}\}$ and $J_{2i}\{\theta,\hat{\nu}_{i};Y_{i}\}$ have the  order $O_{p}({1}/{n_{i}})$.  Therefore, we obtain
\begin{align}
m(\theta;Y) &= \sum_{i=1}^K m_i(\theta;Y_i)=\sum_{i=1}^{K}[h_{i}\{\theta,\hat{\nu}_{i};Y_{i}\}-\frac{1}{2}\sum_{i=1}^{K}\log[-h_{i}^{(2)}\{\theta,\hat{\nu}_{i};Y_{i}\}/2\pi] \nonumber\\
&  +\sum_{i=1}^{K}\log[1-C_{n_{i}}\{\theta,\hat{\nu}_{i}\}]+O_{p}(n^{-1}).\label{marapp}
\end{align}
The first two terms in (\ref{marapp}) are also called the adjusted profile likelihood.
The equation (\ref{marapp}) indicates the following theorem,
\begin{theorem} Under Conditions in Appendix,
the marginal likelihood
$$m(\theta;Y)=p_{\nu}(H)+O_{p}(\frac{1}{n}).$$
\end{theorem}

From Theorem 3, $p_{\nu}(H)$ is a perfect approximation to the marginal log-likelihood $m(\theta;Y)$.
Hence,
we directly give an estimator of $\beta$ by maximizing $p_{\nu}(H)$ with respect to $\beta$ instead of
$m(\theta;Y)$. Let
$\tilde{\beta}$ and $\hat{\beta}$ be maximizers of  $m(\theta;Y)$ and  $p_{\nu}(H)$, respectively. Next theorem shows a connection between $\tilde{\beta}$ and $\hat{\beta}$,
which proof is in Appendix.
\begin{theorem}
Under Conditions in Appendix,  for given $\nu$ and $\sigma^2$, we have $$\tilde{\beta}=\hat{\beta}+O(\frac{1}{n}).$$
\end{theorem}	
 Then we can use $\hat{\beta}$ as an estimator of
$\beta$. To make inference on $\beta$, we need to know the variance of $\hat{\beta}$.
Let
\begin{align}
&V = \left (
 \begin{matrix}
   V_{11} & V_{12}  \\
   V_{21} & V_{22}
  \end{matrix}
  \right )  =n\left(
 \begin{matrix}
   Var(\hat{\beta}) & Cov(\hat{\beta},\hat{\nu}-\nu)  \\
   Cov(\hat{\nu}-\nu,\hat{\beta}) & Var(\hat{\nu})
  \end{matrix}
  \right), \nonumber\\
 & M=\frac{1}{n} \left(
 \begin{matrix}
   -\frac{\partial^2H}{\partial\beta\partial\beta^T}  &  -\frac{\partial^2H}{\partial\beta\partial\nu^T}  \\
    -\frac{\partial^2H}{\partial\nu\partial\beta^T}  &  -\frac{\partial^2H}{\partial\nu\partial\nu^T}
  \end{matrix}
  \right)\Bigg |_{\beta=\hat\beta,\nu=\hat\nu}. \nonumber
 \end{align}
If $E(M)$ is non-singular, under appropriate conditions we show that $M^{-1}$ converges to $V$ as $n \rightarrow \infty$.
Thence, $M^{-1}$ can be used to estimate the variance of $\hat\beta$. We use the result of \cite{lee1996hierarchical} and we can see the proof in their appendix.

\subsection{Estimation of dispersion parameter}

It is well-known that for mixed linear models, in order to reduce bias, a restricted log-likelihood \citep{patterson1971recovery} is used to estimate the dispersion parameters.
For model (\ref{rmodel-obs}), the restricted log-likelihood is
\begin{equation}
r=\log \,L(\sigma^2; Y | \hat{\beta}) \equiv \log \, f_{\sigma^2}(Y | \hat{\beta}).\nonumber
\end{equation}
For mixed linear model, \cite{cox1987parameter} extended $r$ to $p_{\beta}(m)$.
To avoid intractable integration in $p_{\beta}(m)$, following \cite{lee2001hierarchical}
we use $p_{\beta,\nu}(h)$ to approximate $p_{\beta}(m)$.  Maximizing  $p_{\beta,\nu}(h)$ gives an estimate of the dispersion parameter. We know that logarithm transformation of model (\ref{rmodel-obs}) is a mixed linear model such that h-likelihood for the logarithm transformation model differs only a constant (in Jacobi matrix) from $H\{\theta,\nu;Y\}$.
Hence, from \cite{lee2001hierarchical}, maximizing $ p_{\beta,\nu}(h)$ provides a reasonable dispersion estimators. For further details, please see \cite{lee2001hierarchical}.

\subsection{Inference procedure}

From suggestion in \cite{lee2005likelihood}, we generally use the h-loglihood $H$, the marginal likelihood $m$ and the restricted loglihood $p_{\beta}(m)$ for inference of $\nu$, $\beta$ and $\sigma^2$, respectively. Traditionally, we use sampling method like Monte Carlo simulation to calculate m. In our method, to avoid the calculation of integration $m$, we use $p_{\nu}(h)$ and $p_{\nu,\beta}(h)$ to estimate $\beta$ and $\sigma^2$ instead of  $m$ and $p_{\beta}(m)$. Therefore, the estimation equations are
\begin{eqnarray}
\left\{
\begin{array}{lll}
\frac{\partial H}{\partial \nu} = 0\\
\\
\frac{\partial p_{\nu}(H)}{\partial \beta} = 0\\
\\
\frac{\partial p_{\nu,\beta}(H)}{\partial \sigma^2} = 0.\label{eqnest}
\end{array}
\right.
\end{eqnarray}

Iteration algorithm such as Newton-Raphson algorithm is applied to solve the equation (\ref{eqnest}). From our simulation results, the estimation procedure
converges quickly, for example with two or three iterations.
However, when number of repeated measurement $K$ and sample size $n$ are large, computation of the matrices involved in estimation procedure becomes much more complicated.
Under this case, we can compile the subprogram to overcome this shortcoming by using  $C$  or python language.

\subsection{A mixed linear model with known variance of error}

Since the error term $\epsilon$ in model (\ref{rmodel-obs}) have a specific distribution in $(\ref{density})$, we obtain density function of $\log(\epsilon)$
\begin{equation}	
	f(t)=c\exp(-e^t-e^{-t}+2),
\end{equation}	
where $c\approx 0.594$ is a normalizing constant. By Taylor expansion on  $e^t$, it is amazing to find $\log(\epsilon)$ behaves almost like a normal distribution with mean $0$ and standard variance $\phi=0.6434$. So we also compare the proposed method with the following logarithm transformation model
\begin{equation}\label{log-rmodel}
	\tilde{Y}_{ij}=\log(Y_{ij})=X_{ij}^T\beta+\nu_{i}+e_{ij},i=1,....K,j=1,....n,
\end{equation}	
where $e_{ij}$ has normal distribution with mean 0 and standard variance $\phi$.
Let $\tilde{Y}_i=(\tilde{Y}_{i1},...,\tilde{Y}_{in})^T$ and $\tilde{Y}=(\tilde{Y}^T_1,...,\tilde{Y}^T_K)^T$.
It easily shows that $\Sigma=Cov(Y)=\sigma^2(I_{K} \otimes 1_{n}1_{n}')+\phi^2I_{Kn}$,
where $1_{n}$ is a length $n$ vector with all elements 1, $I_{q}$ is an identical matrix with rank $q$, and $\otimes$ stands for Kronecker product.

To fit model (\ref{log-rmodel}),
the least square (LS) method is used to provide equation
\begin{equation}	
	X^T\Sigma^{-1}X\beta=X^T\Sigma^{-1}\tilde Y, \label{linmixest}
\end{equation}	
where $X=(X_{1}^T,...,X_{K}^T)^T$,$X_{i}=(X_{i1},...,X_{in})^T$. Solving (\ref{linmixest}) gives an estimate of $\beta$.

\section{Numerical study}

\subsection{Simulation study}

Simulation studies are constructed to evaluate finite sample performance of the proposed method.
Simulation data are generated independently from model (\ref{rmodel}), where  covariates have uniform distribution on on $(0,1)$ and random effect have standard normal distribution.
 The error term has three distributions: $E_1$, $E_2$ and $E_3$,
where $E_1$ is (\ref{density}), $E_2$ is an exponential normal distribution with mean $0$ and variance $0.414$,
and $E_3$ is an exponential of the uniform distribution on $(-2,2)$. Sample size $(K,~n_i)=(10,~10)$, $(20,~5)$ and $(20,~10)$.
Parameter $\beta^\top=(\alpha,\beta_1,\beta_2,\beta_3)^\top$ $=(2,2,1,1)^\top$.
All of simulation are repeated 500 times.

Following \cite{paik2015frequentist}, we separately consider cases of known and unknown dispersion parameter $\sigma^2$, where
$\sigma^2$ is settled with value 1 for the known case  while needs to be estimated for the unknown one.
Here, performance of the proposed h-likelihood relative error method (HRE) is compared with that of the traditional least square method (LSE) mentioned in subsection 2.6. 
Tables $1$ and $2$ present results of parameter estimation from HRE and LSE for the known and unknown dispersion parameters, respectively.
We can see that all estimates of parameter $\beta$ from HRE and LSE  are very close to their true values. However, under E3, HRE has smaller standard deviation than LSE, while
under E1 and E2 these two methods have comparable results. From Table 2, under E3 estimates of $\sigma^2$ have larger bias for LSE while HRE performs well.
As $K$ or $n_i$ increases, the standard deviations of the parameter estimates become smaller.

\subsection{Application}

The proposed method is applied to two datasets, cakes data and sleepstudy data. The cake data contains $n_i=6$ different baking temperatures ranged from 175$^\text{o}C$ to 225$^\text{o}C$, and three different recipes. Among each recipe, there were $K=15$ replications.  It assumes that these replications have a randomized blocks scheme:
one by one is produced, so that the differences among replicates may represent time effect. The response here is breaking angle, covariate is temperature. Since the breaking angle is gradual, it tends to have a subjective element (random effect). We can find sleepstudy data in package lme4. The average reaction time per day for each subject is recorded in a sleep deprivation study. On day 0 the $K=18$ subjects had their normal amount of sleep. Starting that night they were restricted to 3 hours of sleep per night and were measured $n_{i}=10$ days. The responses are the average reaction time on a series of tests in each day, covariate is date and random effect is brought in by subjective effect.

To evaluate the performance of HRE and LSE, each dataset is partitioned into two parts: around 2/3 samples as training data and the left as test data.
The prediction accuracies from these two methods are measured by four different median indices: median of absolute prediction errors $\{\vert Y_{i}-\hat{Y_{i}} \vert\}$ (MPE), median of product relative predition errors $\{\vert Y_{i}-\hat{Y_{i}} \vert ^2 /Y_{i}\hat{Y_{i}}\}$ (MPPE), median of additive relative prediction errors $\{\vert Y_{i}-\hat{Y_{i}} \vert/Y_{i}+\vert Y_{i}-\hat{Y_{i}} \vert/\hat{Y_{i}}\}$ (MAPE)  and median of squared predition errors $\{\vert Y_{i}-\hat{Y_{i}} \vert ^2 \}$ (MSPE), where $\hat{Y}_{ij}=\exp(X_{ij}^ {\top}\hat{\beta}+\hat{\nu}_{i})$ for HRE.
Parameter estimates and prediction results are shown in Table 3. It shows that HRE and LSE have similar estimates for $\beta$, but HRE has much smaller variance estimate than LSE.
For cakes data, the breaking angle is larger when temperature increases, and for sleepstudy data, the average reaction time with longer studied time is larger.
More importantly, all these 4 prediction indices from HRE are smaller than LSE.


\bibliographystyle{elsarticle-harv}
\bibliography{bib}

\vskip 0.2 in
\section*{Appendix: Proofs of the main results}
\vskip 0.2 in
\setcounter{equation}{0}
\renewcommand{\theequation}{A.\arabic{equation}}

Condition $A_{1}$: $n_{i} \rightarrow \infty$ , $n_{i}/K \rightarrow O_{p}(1)$.

Condition $A_{2}$: $|| \frac{\partial}{\partial \theta} W_{i}^{(1)} \{ \theta,\nu_{i};Y_{i} \} || = O_{p}(1)$.
\vskip 10pt	
\noindent{\bf{Proof of Theorem 1}}

The Laplace approximation defined in (\ref{laplace}) each group $i$ becomes
\begin{equation}
\hat{h}_{i} =\hat{m}_{i}(\theta;Y_i)+\hat{l}_{i}(\theta;\nu_i | Y_i)
\end{equation}
where $h_{i}$ is the h-likelihood ,$m_{i}$ is the marginal likelihood of $Y_i$ and $\hat{l}_{i}(\theta;\nu_i | Y_i)$ is the conditional likelihood of $\nu_{i}$ given $Y_i$.Using the power series expansion we can get
\begin{equation}
exp \,h_{i}=exp \,\hat{h}_{i}\{1+c_{3}(\nu_i-\hat{\nu}_i)^3+c_{4}(\nu_i-\hat{\nu}_i)^4+..\}
\end{equation}
where $c_{3}=\frac{1}{6}\frac{h_{i}^{(3)}(\hat{\nu}_i)}{h_{i}(\hat{\nu}_i)}$ and $c_{4}=\frac{1}{24}\frac{h_{i}^{(4)}(\hat{\nu}_i)}{h_{i}(\hat{\nu}_i)}$.Because we have specific expression and $Y_{ij}>0$,we can get that $c_{3}$ and $c_{4}$ are coefficients with $O_{p}(n_{i})$ order,or even $O_{p}(1)$.Tierney and Kadane(1986) showed that $exp \,m_{i}=exp\,\hat{m}_{i}\{1+O_{p}(n_{i}^{-1})\}$.So we  can show that $l+O_{p}(n_{i}^{-1})=\hat{l}\{1+c_{3}(\nu_i-\hat{\nu}_i)^3+c_{4}(\nu_i-\hat{\nu}_i)^4+..\}$,therefore $l_i=\hat{l}_i\{1+c_{3}(\nu_i-\hat{\nu}_i)^3+c_{4}(\nu_i-\hat{\nu}_i)^4+..O_{p}(n_{i}^{-1})\}$.Because $\hat{l}(\theta;\nu_i | Y_i)$ is the log-likelihood of the normal density,therefore $E(\nu_{i} | Y_i)=\hat{\nu}_{i}+(c_{3}-\hat{\nu}_{i}c_{4})E^*(\nu_{i}-\hat{\nu}_{i})^4=\hat{\nu}_{i}+O_{p}(n_{i}^{-1})$


\vskip 10pt	
\noindent{\bf{Proof of Theorem 4}}

Using the fact that
\begin{equation}
\int f_{\theta}(\nu | y)d\nu=1
\end{equation}
we can get the conclusion that
\begin{equation}
E(\partial h / \partial \theta | y)=\partial m / \partial \theta +E(\partial log\, f_{\theta}(\nu | y) /\partial \theta |y)=\partial m / \partial \theta
\end{equation}
Consider the Taylor series expansion
\begin{equation}
\partial h / \partial \beta_{k} =\partial h /\partial \beta_{k} |_{\nu=\hat{\nu}}+A_{1}(\nu-\hat{\nu})+A_{2}(\nu-\hat{\nu})^2/2!+....,
\end{equation}
where $A_{1}=(\partial / \partial \beta_{k})(\partial h / \partial \nu)|_{\nu=\hat{\nu}}$ and $A_{2}=(\partial / \partial \beta_{k})(\partial^2 h / \partial \nu^2)|_{\nu=\hat{\nu}}$.Since $\hat{\nu}=E(\nu | y)+O_{p}(n^{-1})$ , $var(\nu | y)=O_{p}(n^{-1})$ and $A_{i}=O_{p}(n)$,Equation B.3 becomes $E\{\partial h/\partial \beta_{k} | y\}=\partial h /\partial \beta_{k} |_{\nu=\hat{\nu}}+O_{p}(1)$.Let $\hat{\beta}_{k}$ be the solution of $\frac{\partial m}{\partial \beta_{k}}=0$ and $\tilde{\beta}_{k}$ be the solution of $\frac{\partial p_{\nu}(h)}{\partial \beta_{k}}=0$,Equation A.5 becomes
\begin{equation}
\frac{\partial m}{\partial \beta_{k}}=\frac{\partial h}{\partial \beta_{k} }|_{\nu=\hat{\nu}}+O_{p}(1)
\end{equation}
and we can use Tylor series expansion again and equation B.4 becomes
\begin{align*}
\frac{\partial m}{\partial \hat{\beta}_{k}}&=\frac{\partial h}{\partial     \hat{\beta}_{k} }|_{\nu=\hat{\nu}}+O_{p}(1)\\
&=\frac{\partial h}{\partial     \tilde{\beta}_{k} }|_{\nu=\hat{\nu}}+(\hat{\beta}-\tilde{\beta})\frac{\partial^2 h}{\partial    \tilde{\beta}_{k}^2 }|_{\nu=\hat{\nu}}+...+O_{p}(1)\\
&=0
\end{align*}
Since $\frac{\partial^{(n)} h}{\partial    \tilde{\beta}_{k}^{(n)} }|_{\nu=\hat{\nu}}$ are of $O_{p}(n)$ order,we can get the conlusion that $\hat{\beta}-\tilde{\beta}$ is of $O_{p}(\frac{1}{n})$ order.

\newpage

\begin{table}
\begin{center}
  \centering
  \caption{Results of parameter estimate with 500 replications in the error distributions of $E1$ , $E2$ and $E3$. In this time the dispersion parameter $\theta$ is known. }
  \vskip 10pt
  \tabcolsep=3pt \fontsize{8}{14}\selectfont
  \begin{tabular}{cl cc ccc c}
     \hline
    Error& (K,n) &method& $\hat\beta_{0}$ & $\hat\beta_{1}$& $\hat\beta_{2}$& $\hat\beta_{3}$ \\
     \hline
$E_{1}$ & (10,10) & HRE  &1.996(0.396)$^*$ &1.991(0.239) & 0.990(0.245) &1.008(0.249) \\

          &&LSE&1.996(0.397&1.989(0.237)&0.991(0.245)&1.008(0.251)\\

        & (20,5) & HRE  &1.980(0.329)&2.010(0.261)& 1.002(0.247)&1.004(0.257) \\

          &&LSE&1.980(0.327)&2.010(0.261)&1.003(0.247)&1.004(0.259)&\\

$E_{2}$ & (10,10) & HRE  &2.009(0.396)&2.016(0.251)& 1.003(0.230)&1.004(0.232) \\

          &&LSE&2.008(0.392)&2.016(0.251)&1.004(0.230)&1.004(0.230)&\\

        & (20,5) & HRE  &1.997(0.329)&1.997(0.257)& 0.998(0.247)&1.009(0.253) \\

          &&LSE&1.997(0.330)&1.998(0.257)&0.998(0.247)&1.007(0.251)&\\

$E_{3}$ & (10,10) & HRE  &2.031(0.483)&2.013(0.374)& 0.967(0.387)&0.988(0.381) \\

          &&LSE&2.035(0.519)&2.012(0.417)&0.963(0.434)&0.987(0.437)&\\

        & (20,5) & HRE  &2.014(0.454)&2.007(0.134)& 0.996(0.399)&0.978(0.399)\\

          &&LSE&2.013(0.480)&2.011(0.459)&1.000(0.435)&0.971(0.436)&\\

     \hline
     \multicolumn{8}{l}{$^*$ Standard deviations are in parentheses. }
   \end{tabular}

\end{center}
\end{table}

\begin{table}
  \centering
  \caption{Results of parameter estimate with 500 replications in the error distributions of $E1$ , $E2$ and $E3$. In this time the dispersion parameter $\theta$ is unknown. }
  \vskip 10pt
  \tabcolsep=3pt \fontsize{8}{14}\selectfont
  \begin{tabular}{cl c c ccc c}
     \hline
    Error& (K,$n_{i}$) &method& $\hat\beta_{0}$ & $\hat\beta_{1}$& $\hat\beta_{2}$& $\hat\beta_{3}$ & $\hat{\sigma^2}$ \\
     \hline
$E_{1}$ & (10,10) & HRE &1.971(0.374)&1.990(0.251)& 1.016(0.243)&1.014(0.221)&1.008 \\

          &&LSE&1.971(0.373)&1.989(0.251)&1.016(0.241)&1..014(0.224)&1.127\\

        & (20,5) & HRE &1.988(0.329)&1.994(0.261)& 1.002(0.259)&1.002(0.245)&0.977 \\

  &&LSE&1.989(0.327)&1.994(0.263)&1.002(0.255)&1.001(0.247)&1.125\\

        & (20,10) & HRE &1.988(0.267)&2.008(0.161)& 1.010(0.163)&0.992(0.167)&0.963 \\

  &&LSE&1.988	(0.269)&2.007(0.164)&1.010(0.164)&0.992(0.166)&1.088\\

$E_{2}$ & (10,10) & HRE &1.975(0.383)&2.000(0.228)& 0.996(0.241)&1.012(0.228)&1.000 \\

          &&LSE&1.976(0.383)&1.999(0.228)&0.996(0.239)&1.011(0.226)&1.114\\

        & (20,5) & HRE &1.985(0.345)&2.006(0.263)& 0.987(0.253)&0.996(0.257)&0.979 \\

     &&LSE&1.987(0.344)&2.006(0.265)&0.986(0.251)&0.996(0.257)&1.132\\

             & (20,10) & HRE &2.011(0.269)&1.993(0.168)& 0.996(0.169)&1.004(0.171)&0.990 \\

  &&LSE&2.010(0.269)&1.993(0.166)&0.997(0.167)&1.004(0.171)&1.118\\

$E_{3}$ & (10,10) & HRE  &2.012(0.491)&2.006(0.375)& 1.000(0.363)&0.991(0.397)&1.037 \\

          &&LSE&2.006(0.519)&2.008(0.425)&1.003(0.409)&0.998(0.445)&0.404\\

        & (20,5) & HRE  &2.016(0.428)&1.972(0.392)& 1.004(0.415)&0.984(0.418)&1.119 \\

     &&LSE&2.032(0.449)&1.958(0.422)&0.989(0.443)&0.984(0.446)&0.430\\

             & (20,10) & HRE &2.024(0.326)&1.988(0.264)& 0.992(0.284)&0.999(0.258)&1.085 \\

  &&LSE&2.029(0.354)&1.982(0.304)&0.990(0.323)&0.998(0.293)&0.453\\

     \hline
   \end{tabular}

\end{table}

\begin{table}
  \centering
  \caption{Comparisons of median prediction errors with HRE and LSE for cakes data and sleepsdtudy data}
  \vskip 10pt
  \tabcolsep=3pt \fontsize{8}{14}\selectfont
  \begin{tabular}{cc ccc cccc}
     \hline

    Data& method & $\hat{\beta}_{0}$  & $\hat{\beta}_{1}$&$\hat{\sigma^2}$ &MPE & MPPE & MAPE & MSPE\\
    \hline
    cakes & HRE & 2.251&0.006&0.003&4.7914& 0.0185 & 0.2728 & 23.038\\
    & LSE& 2.247&0.006&1.142& 5.0339&0.0209& 0.2899& 25.353\\
    sleepstudy & HRE & 5.532&0.033&0.012&31.1827& 0.0077 & 0.1758 & 972.44\\
    & LSE&5.532&0.033&0.899 & 31.3578&0.0085& 0.1845& 983.42\\

     \hline
   \end{tabular}

\end{table}

\end{document}